\documentclass[twocolumn,prb,amsmath,amssymb]{revtex4}
\usepackage{graphicx}

\begin{document}
\title{Optical echo in photonic crystals}
\author{A.E.Antipov}\email{antipov@shg.ru}
\author{A.N. Rubtsov}

\affiliation{Department of Physics, Moscow State University,
119992 Moscow, Russia}

\date{\today}

\begin{abstract}
The dynamics of photonic wavepacket in the effective oscillator
potential is studied. The oscillator potential is constructed on a
base of one dimensional photonic crystal with a period of unit
cell adiabatically varied in space. The structure has a locally
equidistant discrete spectrum. This leads to an echo effect, i.e.
the periodical reconstruction of the packet shape. The effect can
be observed in a nonlinear response of the system. Numerical
estimations for porous-silicon based structures are presented for
femtosecond Ti:Sapphire laser pump.
\end{abstract}

\maketitle

 One of the strongest driving forces of the modern optics is a
projection of the solid-state physics concepts on a photonic
language. Except the fundamental interest, this is related to
so-called photonics: a construction of optical based elements of
information processing devices. Modern technology provides an
opportunity to construct a photonic devices of a sunmicron and micron sizes. In
particular it is worth mentioning photonic crystals \cite{Joan, Yabl}, structures
with a periodic modulation of optical properties. They act on a photon the
same way as a crystal lattice potential on electrons. Beauty
optical analogs of several solid-state phenomena have been
observed in systems, based on photonic crystals. One could mention
Bloch oscillations \cite{BlochOsc0,BlochOsc1}, optical analog of Franz-Keldysh effect \cite{FrKeld0, FrKeld1},
fabrication of optical molecule \cite{Molec}, etc. Optical field localization in the photonic crystal
based structures provides numerous nonlinear optical phenomena \cite{PC0,PC1}.

A description of experimental results in this field commonly uses
spectral(plane-wave) representation. At the same time, modern
femtosecond lasers provide pulses of a small longitude. For
example 100 $fs$ impulse has a $30 \mu m$ length in vacuum and is
even shorter in a material media. This is comparable to the size
of a photonic crystal structure. By taking into account a finite
length of the photonic wavepacket, one can predict a new range of
optical effects, similar to the ones, based on electron wavepacket
spatial localization. In this paper we discuss possible
observation of the ``optical echo'' effect, i.e. periodical
reconstruction of a shape of the optical pulse in a specially
designed photonic crystal structure.

The simplest echo effect can be realized for a quantum particle in
a harmonic oscillator potential. Let the particle being localized at the initial moment near certain spatial point $x_0$. It can be
characterized by the wavefunction $\Psi_0(x)$, for example a
Gaussian packet can be considered $\Psi_0(x)\propto \exp \left(
{-\frac{(x-x_0)^2}{2 R^2_0}+ikx}\right)$. After the initial moment
a space spreading of the pulse takes the place because of the
dispersion. However, after the time equal to the period of the
oscillator $T$, the wavepacket will reconstruct it's shape back.
Indeed, all the eigenfunctions of oscillator evolve with multiple
frequencies of the oscillator, so that the wavefunction of the
packet $\Psi(x,t)$ repeats itself with a period $T$:
\begin{equation}\label{2}\Psi(x,t)=\sum_{n}^{}{C_n\phi_n(x)\exp{(i\frac{2\pi}{T}nt)}}\end{equation}
One can see from (\ref{2}) that the crucial circumstance here
is that eigenlevels of the harmonic oscillator are equidistant in
frequency domain. On the other hand energy levels of almost any
pendulum are equally spaced in energy in the quasiclassical
limit\cite{PVE}. In this case energy split in in neighboring levels is
just $\frac{2\pi\hbar}{T(E)}$, where $T(E)$ is a period of
oscillation at given energy. This means the echo effect can be
observed for packets composed of locally equidistant states of the
discrete spectrum of almost arbitrary potential. Echo effects can
also be observed in more sophisticated cases, for example in the
disordered structures \cite{Pepper}. The only requirement is the
local equidistance of the spectrum.

Our goal is to construct an effective oscillator potential for
optical pulses. This requires a modifying of dispersion relation
with respect to spatial coordinate. Fabricating a
structure with a refractive index being changed smoothly in a wide
range is a very complicated task. However, the
structures based on one-dimensional photonic crystals can be used.
In order to form a structure with a discrete spectrum (i.e.
localized eigenstates), we propose using a photonic crystal with a unit cell
period adiabatically varied in space. Let us imagine that for the
center of the structure the carrier frequency of the wavepacket
lies just above the photonic bandgap. Consider the structure with
a period decreasing to the spatial edges. In this case the photonic
bandgap of a crystal is higher at the crystal edges, than in the geometric center. This
means the wavepacket is reflected from the Bragg mirrors near the
structure borders, therefore it becomes localized in the whole
crystal.

The one-dimensional photonic crystals based on porous silicon are
quite widespread in modern photonic technology. The procedure for
their preparation is quite simple. Such photonic crystals
constitute of repeating pair of layers of $n\simeq 1.5$ and
$n'\simeq 2.2$ refraction indexes. The optical lengths of both layers in each
repeating pair of the crystal are usually taken equal, their material dispersion can be examined as linear.
We will address to consider such type of structures. The adiabatic variation
of the period of such crystal is also feasible technological task.
However in order to obtain sufficient number of equidistant
localized levels (around 10) the whole number of pairs must be
taken large enough. In our model $M=100$ pairs of layers are
considered.

In the effective local linear susceptibility approximation all
layers can be characterized by two values: $d_m$ - the length of
the layer and $n_m$ - it's refraction index. The field strength of
eigenmode of each layer is given by expression:
\begin{equation}\label{eigen}{\cal E}_m(x)=A_me^{-ik_mx}+B_me^{ik_mx}\\
\notag k_m=\frac{w}{c}n_m,\end{equation} where $A_m$ and $B_m$ are
unknown amplitudes, which are determined by frequency of eigenmode
$w$.

There are different ways to calculate the spectrum of the
structure. One possibility would be a direct search of a solution
of eigenvalue problem.  The boundary conditions on the edges of
the structure and on the joints of the layers produce a set of
equations for $A_m, B_m$ for each layer. The solvability of this
set requires finding a determinant of a $2M\times 2M$ matrix with
some elements equal to zero. This would give an equation for $w$.
The numerical solving of this equation produces a spectrum of
whole system.

% \begin{center}
% %\includegraphics[width=0.52\textwidth,height=0.4\textwidth]{pics/eq1.pdf}
% \includegraphics[width=8.5cm]{pics/table1.pdf}
%  %f1.eps: 6815744x0 pixel, 300dpi, 57706.63x0.00 cm, bb=14 14 578 307
% \end{center}

Another possible way to find a spectrum of the system is to treat
it as an effective adiabatically varied periodic potential. This
method is quite similar to employing Vl\"{o}ke or Bloch's theorem. It
fits the paradigm of applying condensed matter physics ideas on
the field of optical problems. Let's zoom on pair of two layers :
$(d,n)$ and $(D-d,n')$. Their overall spatial length is $D$. Each
eigenmode of the pair $j$ can be presented the same way as (\ref{eigen}):
\begin{widetext}
\begin{equation}
{\cal E}_j(x)=A_je^{ikx\Theta(d-x)}+B_je^{-ikx\Theta(d-x)}+
+{A^{\prime}_j}e^{ik'x\Theta(x-d)}+B^{\prime}_je^{-ik'x\Theta(x-d)}
\end{equation}
\end{widetext}
$$ \notag k'=k\frac{n'}{n} $$
 $\Theta(x)$ is the Heavyside function.
 The nil of the coordinates is placed at the beginning of the first layer.
 Similarly to Vl\"{o}ke's theorem we propose that: \begin{equation}
 {\cal E}(x+D)={\cal E}(x)e^{i\chi}
\end{equation}
Let $\Gamma=\frac{n}{n'}$, $k=\frac{w}{c}$, and
$\chi_0=kdn=k(D-d)n'$. Using the boundary conditions on the layer
connections the value of Bloch's phase $\chi$ can be found
\cite{Suhorukov}:
\begin{equation}
 \cos{\chi}=\frac{(\Gamma+1)^2}{4\Gamma}\cos{2\chi_0}-\frac{(\Gamma-1)^2}{4\Gamma}
\end{equation}
That is equivalent to
\begin{equation}\label{blochph}
 \sin{\frac{\chi}{2}}=\frac{\Gamma+1}{2\sqrt{\Gamma}}\sin{\chi_0}
\end{equation}
Actually, \textit{The Vl\"{o}ke theorem itself is not used}, we
just use the corresponding notation. In fact, the phase $\chi$ is
determined only by the parameters of local pair of layers, i.e. it
is solely connected with the pair number. The expression for the
value of the electric field in the pair number $j$ can now be
rewritten as:
\begin{equation}
\label{cmplxf}
 {\cal E}(x)={\cal E}_0(x)e^{i\Sigma_j}
\end{equation}
$$\notag \Sigma_j=\sum_{i=0}^{j}{\chi_i}$$
${\cal E}_0(x)$ is an expression for the spatial part of electric
field in the first pair of layers. For example in odd layers
(\ref{Suhorukov}) will appear as
\begin{equation}\label{fullfield}
 {\cal E}_j(x)=(A_0(k,d_0)e^{ikx}+B_0(k,d_0)e^{-ikx})e^{i\Sigma_j}
\end{equation}
\begin{figure}%[t]
%\vspace{2cm}
% \begin{centering}
  \includegraphics[width=\columnwidth]{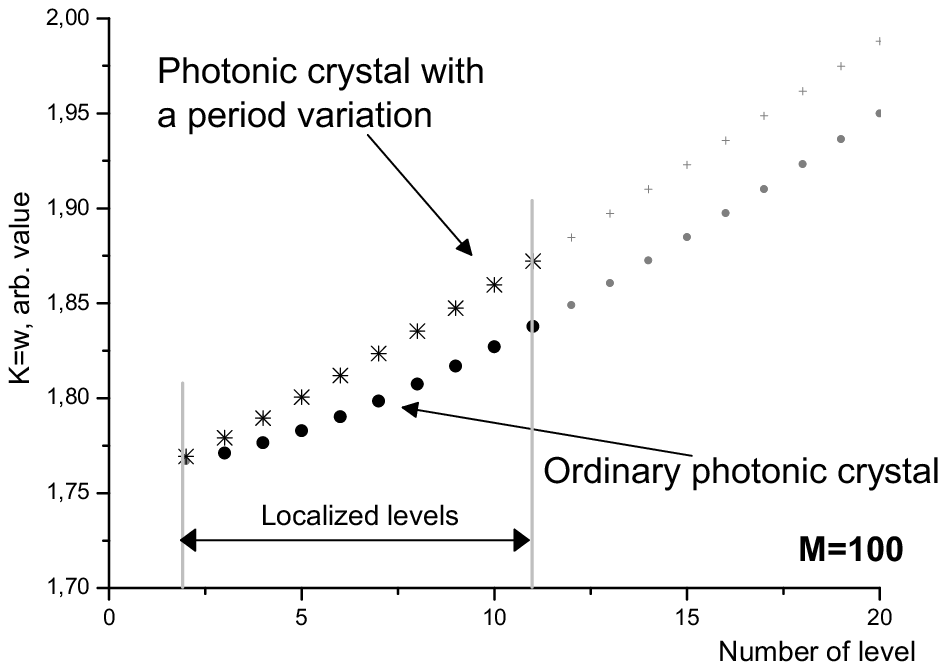}
  \caption{The part of the spectrum of ordinary photonic crystal and a photonic crystal oscillator with 10$\%$ period width variation starting from the upper edge of photonic bandgap spectrum plotted in arbitrary $K=w,\, c=1$ versus the number of the level. Both structures consists of 100 pairs of layers. The important localized levels are highlighted. }
  \includegraphics[width=\columnwidth]{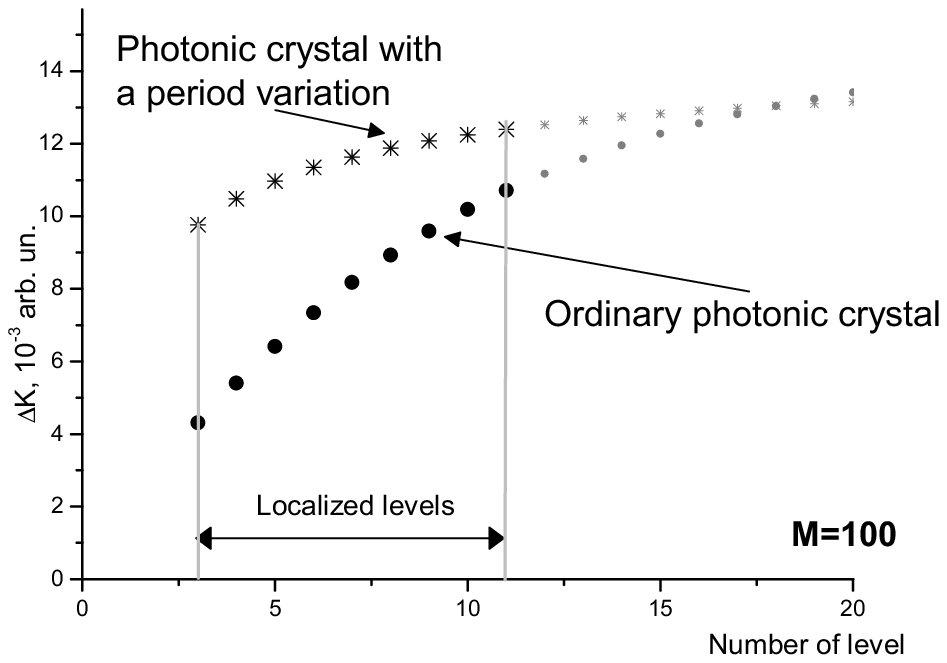}
  \caption{The difference between energy levels versus the number of the level plotted in the same region for same structures, as in Fig. 1. In the frequency interval of the laser signal the photonic crystal with a period variation produces ``more equidistant'' energy levels than an ordinary crystal.}
 %\end{centering}
\end{figure}
\begin{figure}%[!t]
% \begin{centering}
  \includegraphics[width=0.8\columnwidth]{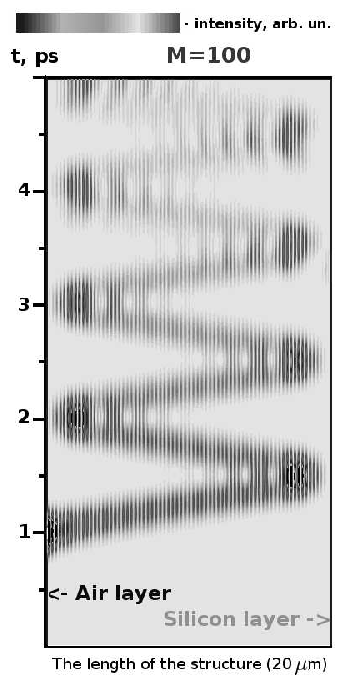}
  \caption{The map of the dynamics of distribution of a part of Gaussian wavepacket with a period of $130$ $fs$ and a $800$ $nm$ wavelength in a photonic crystal oscillator with 100 pairs of layers and a $10\%$ period variation. The spatial coordinate of system is taken over X axis, time is shown on Y axis and intensity of a signal is shown by color.}
% \end{centering}
\end{figure}
The same equations can be derived for complex conjugate ${\cal
E}_j^{*}(x)$. We will consider full field strength as a sum of the
${\cal E}_j(x)$ and ${\cal E}_j^{*}(x)$ which is in fact
$2\Re{{\cal E}_j(x)}$. In this case (\ref{fullfield}) becomes
\begin{widetext}
\begin{equation}
{\cal E}(x)=
(A_0(k,d_0)e^{ikx}+B_0(k,d_0)e^{-ikx})e^{i\Sigma_j}+(A_0^{*}(k,d_0)e^{-ikx}+B_0^{*}(k,d_0)e^{ikx})e^{-i\Sigma_j}
\end{equation}
\end{widetext} In order to get a spectrum of the eigenmodes of this
system it is virtually placed in an opaque resonator. Let $L$ be a
whole length of the system. The boundary conditions on the spatial
edges of the system are
$${\cal E}(0)=0, {\cal E}(L)=0$$
Overall, this gives an equation
\begin{equation}
 \sin{\Sigma}_{M-1}=0,\, \Sigma=\sum_{i=0}^{M-1}{\chi_i}
\end{equation}

The results of the calculation of the spectrum for ordinary
one-dimensional photonic crystal and for one with a period
variation are presented in Fig. 1,2. The localized levels between
the upper bands of photonic bandgaps at the crystal edges and
 in the center of the structure are not exactly equidistant. This means that exact reconstruction of the form of the signal won't occur and the packet will irrevocably lose it's shape after some periods. However this fact is not crucial. Let us introduce the ``non-equidistance parameter'' $\eta=\frac{\delta K}{\Delta K}$, where $\Delta K$ is an effective distance between levels and $\delta k$
is an average deviation of the level interval from the $\Delta K$. For an ideal oscillator $\eta=0\%$, in case of ordinary photonic crystal $\eta=10\%$, while in presence of adiabatic period variation $\eta=2.5\%$. This means in a simple crystal the packet will lose it's shape approximately after 3 periods, while for the selected structure this will happen after 10 periods. This is enough for experimental purposes.

One can consider a possible experimental realization concerning
distribution of a wavepacket inside an observed structure. In
common optical experiments, the sample is surrounded by air and
placed on a substrate. So additional wide layers of air ($n=1$)
and silicon ($n=3.5$) are added on the bounds of the system. We
take a Gaussian shaped wavepacket of a period of $130\:fs$ and the
wavelength $800\:nm$. These are common values for the state of the
art Ti:Sap lasers. The optical lengths of the layers on the
crystal edge are $\frac{\Lambda}{4}=200\:nm$. In this case the
spectrum of the wavepacket lies inside the area of $\sim 10$
localized levels. The distribution of a signal is calculated in
accordance of Eq.(\ref{2}). Fig. 3 shows the results of
calculation. After the initial moment, most of the pulse is
reflected from the structure, because its carrier frequency
corresponds to the photonic bandgap of crystal edge region.
However, approximately $30\%$ part of the signal in terms of
intensity penetrates inside the oscillator due to the tunnelling
and becomes localized. It  starts a propagation with a periodical
shape reconstruction. The localization due to the reflection of
the wavepacket from the effective Bragg mirrors on the edges of
the crystal leads to a periodical stops and changing of direction
of the pulse propagation. The period of echo motion in thus
created optical oscillator potential is $1\:ps$.

Finally, let us discuss how the proposed optical echo effect can be observed in
photonic crystal oscillator. One of the ways to detect the shape reconstruction is a time-resolved observation of
the nonlinear response of the system. Indeed, near the stop-points
of the oscillatoric motion the pulse group velocity falls. Therefore, the
pulse spatial length is smaller and its electromagnetic field is
larger near these points, as can be observed from Fig. 3.
Consequently, second or third optical harmonic response of the
structure should look like a consequence of sharp peaks with a
sub-picosecond period. This should be seen in the autocorrelation
properties of the nonlinear signal. The third harmonic seems more perspective, as the
effect in higher harmonics is more pronounced. On the other hand,
there maybe an experimental problem concerning the absorbtion of
the third-harmonic signal inside the Si-based structure at 800 nm
pump \cite{2003APS, Porcelaine}. This may require usage of
the infrared pump laser (Ti-sapphire + parametric oscillator
system).

Authors are grateful to O.A. Aktsipetrov for his valuable
comments. The work was supported by ``Dynasty'' foundation.

\end{document}